\documentstyle[12pt,aaspp4]{article}
\tightenlines

\def\lcf{lightcone fluctuation}

\begin{document}
\vskip -.8 true in 
\title{
Quantum Gravity and Astrophysics: The Microwave Background
and Other Thermal Sources}   

\author{Rosanne Di\thinspace Stefano}
\affil{Harvard-Smithsonian Center for Astrophysics, Cambridge, MA 02138}
\affil{Department of Physics and Astronomy, Tufts University, Medford, MA 02155}

\author{L.H. Ford}
\affil{Department of Physics and Astronomy, Tufts University, Medford, MA 02155}

\author{Hongwei Yu}
\affil{Institute of Physics and Department of Physics, 
        Hunan Normal University, 
        Changsha, Hunan 410081, China}
\affil{Department of Physics and Astronomy, Tufts University, Medford, MA 02155}
\author{D.J. Fixsen}
\affil{Laboratory for Astronomy and Solar Physics, 
Code 685, NASA GSFC, Greenbelt, MD 20771} 
\begin{abstract}

\vspace{-.2 true in} 
The problem of formulating a fully consistent quantum theory of gravity has
not yet been solved. 
Even before we are able to work out the details of
a complete theory, however, we do know some important qualitative
features to be expected in any quantum theory. 
Fluctuations of the metric, for example, are expected and are associated
with fluctuations of the lightcone.   
Lightcone fluctuations affect the arrival time of signals from distant
sources in potentially measurable ways, broadening the spectra. 
In the work described here, we start with a thermal spectrum and derive the
form of
 spectral changes expected
in a wide class of quantum theories of gravity. 
The results can be applied to any
thermal spectrum.  We focus on their application to the
cosmic microwave background (CMB) because the CMB offers two advantages: 
(1) deviations from a thermal
spectrum are well constrained, and (2) the radiation emanates from
the most distant source of light, the
surface of last scattering.
We use the existing CMB data to derive an upper bound on
the value of $\Delta\, t,$ the mean spread in arrival times
due to metric fluctuations: 
${\Delta t} < 2.1 \times 10^{-14}$ s at the $95\%$ confidence limit. 
This limit applies to a wide range of quantum theories of gravity,
and thus serves to falsify theories predicting a larger
spread in arrival times. We 
find this  limit rules out at least one quantum gravity theory,
the 5-dimensional quantum theory in which the 
``extra" dimension is flat. Meaningful tests of many other models
may also be possible, depending on the results of calculations
to predict 
values of $\Delta\, t,$ and also a second
time scale $\tau_c,$ the correlation time, which is the characteristic
time scale of the metric fluctuations.   
We show that 
  stronger limits on the value of $\Delta\, t,$  and hence on
the effects of \lcf s, are likely to be
derived through
observations of higher-temperature sources, especially in the
X-ray and gamma-ray regimes.
\end{abstract}
\vskip -0.4 true in 
\vskip -0.3 true in

\section{Introduction}

Because the theory of gravity describes the spacetime itself,
quantization of gravity is conceptually more difficult than
quantization of the other fundamental interactions.
If, however, the basic quantum principles we are already
familiar with apply as well to a quantum theory
of gravity, we can make some predictions about
expected quantum effects, even in the absence of a fundamental 
underlying theory. 
Indeed, if some such predictions can be tested,
we may be able to derive useful constraints on the properties
of the true underlying theory in which gravity is quantized.
The basic question we pose here is whether 
data
received from astrophysical systems emitting thermal  radiation
can provide 
tests of quantum gravity.

One generic prediction is that the quantization of gravity
should be associated with fluctuations of the spacetime metric.
Let the classical spacetime metric be represented by 
$\bar g_{\mu \nu} = \eta_{\mu \nu}$, and let the
fluctuations be described
by $h_{\mu \nu}.$ The full quantum metric, $g_{\mu \nu}$, is
therefore the sum: $g_{\mu \nu} = \eta_{\mu \nu}+h_{\mu \nu}$.    
The fluctuations described by $h_{\mu \nu}$ can be viewed as leading to 
fluctuations
of the lightcone. Lightcone fluctuations
cause the arrival times of light signals
to differ from what they
would have been in the classical theory. The arrival times are
not systematically earlier or later than those predicted classically,
but are instead spread about the classical prediction in both directions.
See Figure 1.
The functional form of the mean spread, $\Delta t$, is determined by
the functional form of $h_{\mu \nu}$.
Because increasing path length provides more opportunities for
\lcf s to act, 
$\Delta t$ typically grows with increasing distance 
between the emitter and detector, but the rate of growth can also depend on the geometry and topology of the spacetime.  

\begin{figure}[hbtp]
\leavevmode\epsfxsize=5.6in
\epsfbox{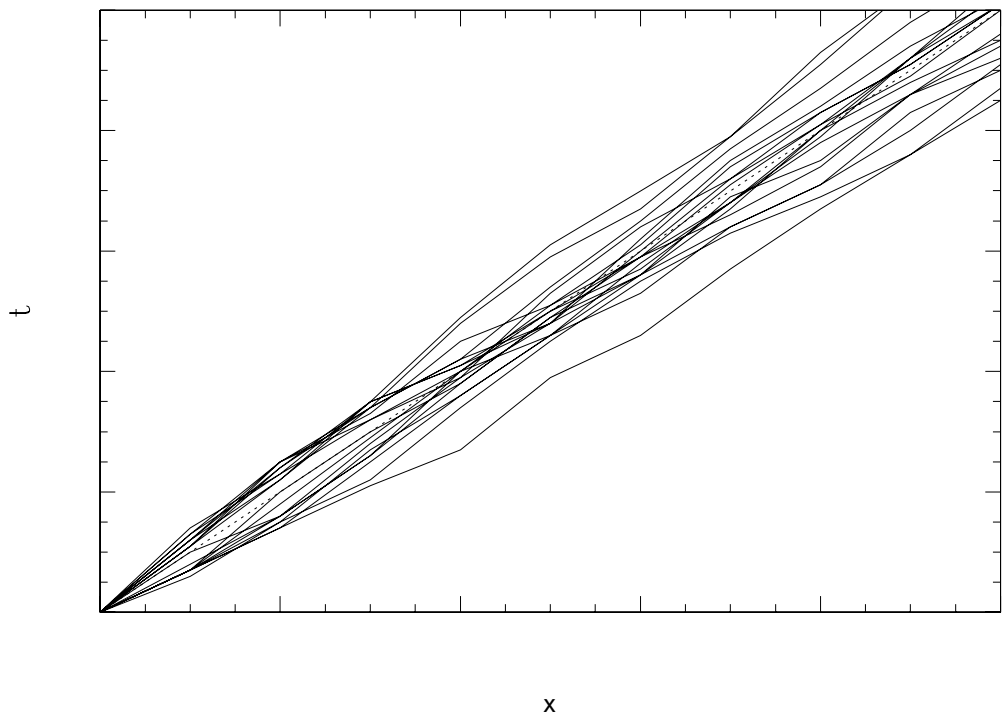}
\vspace{-2.5 true in}
\caption{
Imagine  several photons emitted from the same spacetime point.  
The effects of  lightcone fluctuations on the paths of these photons
 are illustrated by plotting the
time $t$ taken for light to propagate along a path 
coordinatized by $x$; the units are arbitrary.
The dotted line shows the classical 
lightcone; eighteen paths subject to lightcone fluctuations
are depicted by solid lines. Fluctuations
which could
either increase or decrease the time of flight were 
 randomly introduced along each path.
Note that the mean fluctuation,
$\Delta t,$ tends to grow as the path length 
increases.
}
\label{fig=fig1}
\end{figure}

A spread in arrival times introduces spectral changes. If, for example,
the classical situation is that a perfect wave of one frequency
is emitted by some source, lightcone fluctuations will
introduce a spread in the arrival times of each peak.
This spread is measured as a spread in the frequencies received
from the source by a detector. 
Spectral changes between the emitted and received spectrum of light
can be expressed in terms of $\Delta t$.

The purpose of this paper is to compute the form of the changes induced
in a thermal spectrum and to consider the astrophysical circumstances under 
which these changes might be detectable. We focus on the application
to the cosmic microwave background. Our motivation for this application is
twofold. First, deviations of the CMB from a pure thermal spectrum
are well constrained; second, the effect of any given metric fluctuation
increases with distance, and the surface of last scattering is the most distant
emitter of light.  In \S 2 we  start with a given $\Delta t$ and compute 
the form of the expected spectral deviations. 
In \S 3 we 
apply our results to the microwave background.
In \S 4 we demonstrate how $\Delta t$ can be related to metric fluctuations and
thus, how either measuring or placing limits on the size of $\Delta t$
can help to probe quantum fluctuations of the metric. 
In \S 5 we study the prospects that astrophysical measurements, including but not
limited to measurements of the CMB, will allow us to place useful constraints in the
foreseeable future.

It is interesting to consider whether angular dependence due to metric fluctuations,
provides another effect which
   can be used to test quantum gravity. This question is not central to the work we describe
on the cocmic microwave background (CMBR), since we are interested in the
   in the determination of its temperature, and not in the variation
   of temperature across the sky. 
Nevertheless, angular effects could be useful in other
applications. We therefore consider them in the appendix. 
We show that, in the quantum gravity theory we have considered,
 angular deviations are a higher
order effect than lightcone fluctuations and can therefore be
neglected. In other cases, however, 
an angular spread due to \lcf s could turn out to be important; we therefore discuss the implications in \S 5.1.

\section{Deviations from a Thermal Spectrum Due to Light Cone Fluctuations}

A classical pulse of light will propagate through a distance $r$ in
time $r/c=r,$ where, in this last equality, we have introduced units with $c=1.$   
Lightcone fluctuations cause propagation times to be
either longer or shorter than $r$ by a
mean deviation $\Delta t.$ 
\begin{equation}
\Delta t = {{\sqrt{\langle\sigma_1^2\rangle}}\over{r}},
\end{equation}
where ${\langle\sigma_1^2\rangle}$  
is the mean square deviation in the geodesic interval function.

The functional form of $\Delta \, t$ and the range of its expected
values emerge from specific quantum theories.
In general, $\langle \sigma_1^2 \rangle$ is a function of the flight time
$r$. If the fluctuations were a random walk type of process, one could
think of them as occuring in a sequence of steps, each of which is
uncorrelated with the previous step. In this case, one would have
$\langle \sigma_1^2 \rangle \propto r^2$ and $\Delta t \propto \sqrt{r}$.
However, quantum metric fluctuations are not necessarily of this type,
and detailed models can lead to other functional dependencies of
$\Delta t$ upon $r$. One class of models which can lead to large
lightcone fluctuations are Kaluza-Klein type theories with compact
extra dimensions. Confining a quantum field into a region with a
small compact spatial dimension leads to enhanced quantum fluctuations,
as a consequence of the uncertainty principle. This issue, and the
details of how $\langle \sigma_1^2 \rangle$ is calculated will be
discussed in more detail in \S 4. Another issue also discussed there
is that of the correlation time, $\tau_c$. The expected relative
time delay or advance of a pair of pulses is only given by $\Delta t$
when the pulses are separated in time by an amount greater than
$\tau_c$, which one can regard as the characteristic time scale for the
metric fluctuations.                                    

 Here we assume Gaussian fluctuations and that $\Delta t$ is
frequency independent. This is a natural choice, 
and
corresponds to the achromaticity of gravitational effects
in classical general relativity.
The choices of frequency-independence and Gaussianity
are more specifically motivated by work (Ford 1995) showing
that linear quantum metric fluctuations
due to a bath of gravitons modifies the retarded Green's function
from a delta function on the forward lightcone into a Gaussian function.
In addition,
the same behavior was found by Yu and Ford (1999) in models in which
the metric fluctuations arise from compact extra dimensions. In both
of these cases, the resulting $\Delta t$ is found to be
frequency-independent.
Thus, although the assumptions of frequency-dependence and Gaussianity
are not universally applicable, they do apply to important classes
of theories and
are likely to provide useful insights into the effects of
lightcone fluctuations in other classes of quantum gravity theories.
We further note that, although
 some quantum gravity theories
do predict frequency dependent effects (Amelino-Camelia {\it et al.} 1998),
astronomical data is already beginning to place
limits on such
frequency-dependence.
For example, observations of the Crab pulsar rule out differential
time delays greater than $0.35$ msec for energies above $2$ GeV
as compared with $70-100$ MeV (Kaaret 1999). Study of a TeV
$\gamma$-ray flare associated with the active galaxy Markarian 421
indicates that there is no significant time delay between the emission
at $1$ TeV and the emission at $2$ TeV (Biller {\it et al.} 1999).
Multiwavelength observations of $\gamma$-ray bursts can place
even stronger
limits on the frequency-dependence of lightcone fluctuations
if simultaneous observation of the arrival times from radio to $\gamma$-ray
energies can be made, and if the distance to the burst source can be determined
(Schaefer 1999).
We therefore proceed with calculations incorporating the
choices of frequency-independence and Gaussianity, which
simplify the calculations and also allow us to
determine what functional forms for the frequency dependence
can be most readily tested.

A spread in flight times for pulses necessarily implies a
spread in frequency. Consider, for example, a source which emits
a monchromatic wave. The individual wave crests may be regarded
as being pulses whose times of arrival at a detector vary by
about $\Delta t$. This introduces a spread in frequency, $\Delta \nu$.
This means that we can look for signatures of lightcone fluctuations
in the frequency domain rather than the time domain, which is likely to
lead to more stingent constraints. Even a staionary source such a
blackbody emitter, which does not have any intrinsic time dependence,
can have its frequency spectrum distorted by lightcone fluctuations.

When there are Gaussian lightcone fluctuations, 
a monochromatic wave of frequency $\nu_0 = 1/T_0,$ with $T_0$ the period,
becomes a Gaussian with width 
\begin{equation}
|\Delta\, \nu| = {{\Delta\, t}\over{T_0^2}} = \nu_0^2\, \Delta\, t  
\end{equation}
That is,
\begin{equation}
\delta(\nu_0)\longrightarrow 
\exp\Bigg[{-\Big({{\nu-\nu_0}\over{\Delta\, \nu}}\Big)^2}\Bigg] 
{{1}\over{\sqrt{\pi} \Delta\, \nu}}   
\end{equation}
Thus, a spectrum described by a function $F_0(\nu')$, is distorted and
will now be described by $F(\nu).$ If the original spectrum was thermal,
then 
\begin{equation}
F_0(\nu')={{a\, \nu'^3}\over{e^{b\, \nu'}-1}}
\end{equation}
and
\begin{equation}
F(\nu)= { {a} \over{ \sqrt{\pi}\, \Delta\, t} } \int_0^{\infty} d\nu'
{{\nu'}\over{e^{b\, \nu'}-1}}\,  
\exp\Bigg[{-\Big({{(\nu'-\nu)}\over{\nu'^2 \Delta\, t}}\Big)^2}\Bigg].  
\end{equation}
It is convenient to evaluate this integral by introducing the
variable $z,$
defined so that
\begin{equation}
z\, \Delta t = \nu'-\nu.
\end{equation}
$F(\nu)$ can now be expressed as 
\begin{equation}
F(\nu)= { {a} \over{ \sqrt{\pi} }}\int_{-{{\nu}\over{\Delta\, t}}}^\infty
dz\, e^{-{{z^2}\over{\nu^4}}} {{\nu}\over{e^{b\, \nu}-1}}.
\end{equation}
Expanding to second order in $\Delta\, t,$ we find that
\begin{equation}
F(\nu)=F_0(\nu)\, \Bigg[ 1 + f_2(\nu) \Bigg], 
\end{equation}
where $f_2(\nu)$ may be written as follows:   
\begin{equation}   
f_2(\nu)=f_2(x[\nu])=\Bigg({{\Delta\, t}\over{b}}\Bigg)^2 
\Bigg\{ {{x^2}\over{4\, (e^x-1)^2}}\Bigg[e^{2\, x} \Big(x\, (x-14)+42\Big)
+ e^x \Big(x\, (x+14)-84\Big) + 42  
\Bigg]\Bigg\},
\end{equation}
with
\begin{equation} 
x=x(\nu)=b\, \nu \,.
\end{equation}     

The physical meaning of $f_2(x)$  is that it is the 
fractional deviation from a thermal spectrum 
caused by lightcone fluctuations.
It is useful to concentrate on its functional form,
independent of the value of $\Delta \, t$ and of the temperature of
the emitter. We therefore consider $\hat{f_2}(x).$
\begin{equation}
 \hat{f_2}(x) = f_2(x)\ \Big({{b}\over{\Delta\, t}}\Big)^2 \,.
\end{equation}
Figure 2 displays the functional form of $\hat{f_2}(x)$ for $x<10.$ 
As $x$ increases, $\hat{f_2}$ continues to increase as $x^4/4.$
Thus, reliable measurements of a thermal spectrum that extend to
high frequencies with small uncertainties have a good chance of detecting
any deviations that may exist due to lightcone fluctuations. Since, however,
the spectrum is actually falling to zero as $\nu$ increases,
high precision measurements may be more difficult. It is
therefore instructive to consider $\hat{F_2}=F_0\, \hat{f_2},$ which is
plotted in Figure 3. The term that must be
added to the original spectrum in order to obtain the perturbed
spectrum is simply $\hat{F_2} \times ({{\Delta\, t}\over{b}})^2.$ 

The effects of \lcf s on the spectrum can be seen directly in Figures 4 and 5.
In Figure 4, $\Delta t/b$ was chosen to be $0.1.$ 
Note that, as the functional form
of $\hat {f_2}$ dictates, the flux is enhanced near and below the peak of the
thermal spectrum, with a dip at intermediate frequencies, and then 
a second significant enhancement extending out to higher frequencies.
Had the value of $\Delta t/b$ been slightly larger, the flux near $x\sim 8$
would have dipped below $0,$ indicating that the second order approximation
we made to derive a simple analytic form is no longer valid. In Figure 5,
$\Delta t/b$ was chosen to be $0.01.$ With this choice, the effects at
low and intermediate frequency are just visible by eye on a plot linear
in the flux,
while the 
effects at higher frequencies can be clearly seen in a logarithmic
plot. As $\Delta t/b$ decreases further, sensitive measurements
are needed to discover definitive evidence of \lcf s. 
In Figure 6,
the logarithm of the fractional deviation from a thermal spectrum
(expressed in parts per million) 
is plotted as a function of $x.$ Each curve corresponds to a 
different value of $\Delta\, t/b$. This plot illustrates the 
role high-precision measurements of a thermal spectrum can play in
measuring or placing limits on the value of $\Delta t.$

\begin{figure}[hbtp]
\leavevmode\epsfxsize=5.6in
\epsfbox{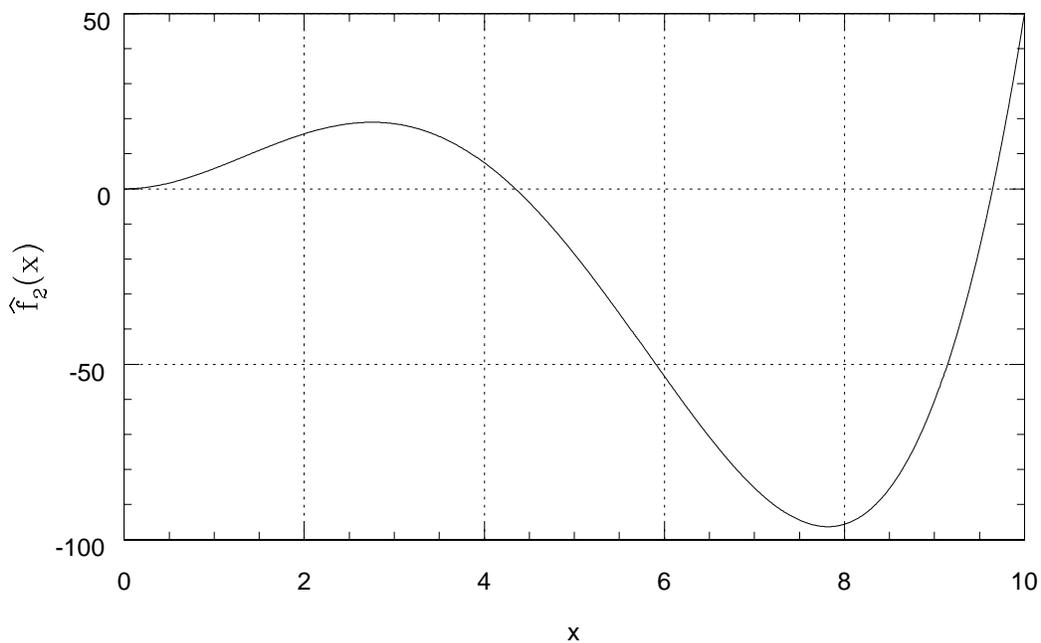}
\vspace{-3.95 true in}
\caption{$\hat{f_2}$ is plotted as a function of $x$.
The mean fractional deviation from a pure thermal spectrum is given
by $\hat{f_2} \times ({{\Delta\, t}\over{b}})^2$.}
\label{fig=fig1}
\end{figure}
\begin{figure}[hbtp]
\leavevmode\epsfxsize=5.6in\epsfbox{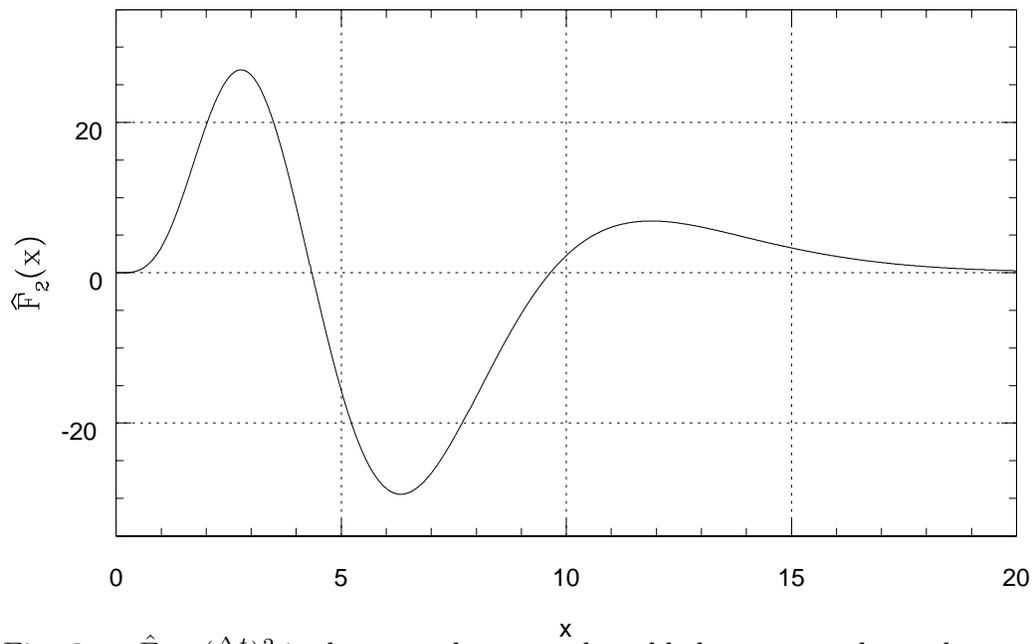}
\vspace{-4.0 true in} 
\caption{$\hat{F_2} \times ({{\Delta\, t}\over{b}})^2$ is the term that must be added to a 
pure thermal spectrum when there are lightcone fluctuations.}
\label{fig=fig2}
\end{figure}
\begin{figure}[hbtp]
\leavevmode\epsfxsize=5.6in\epsfbox{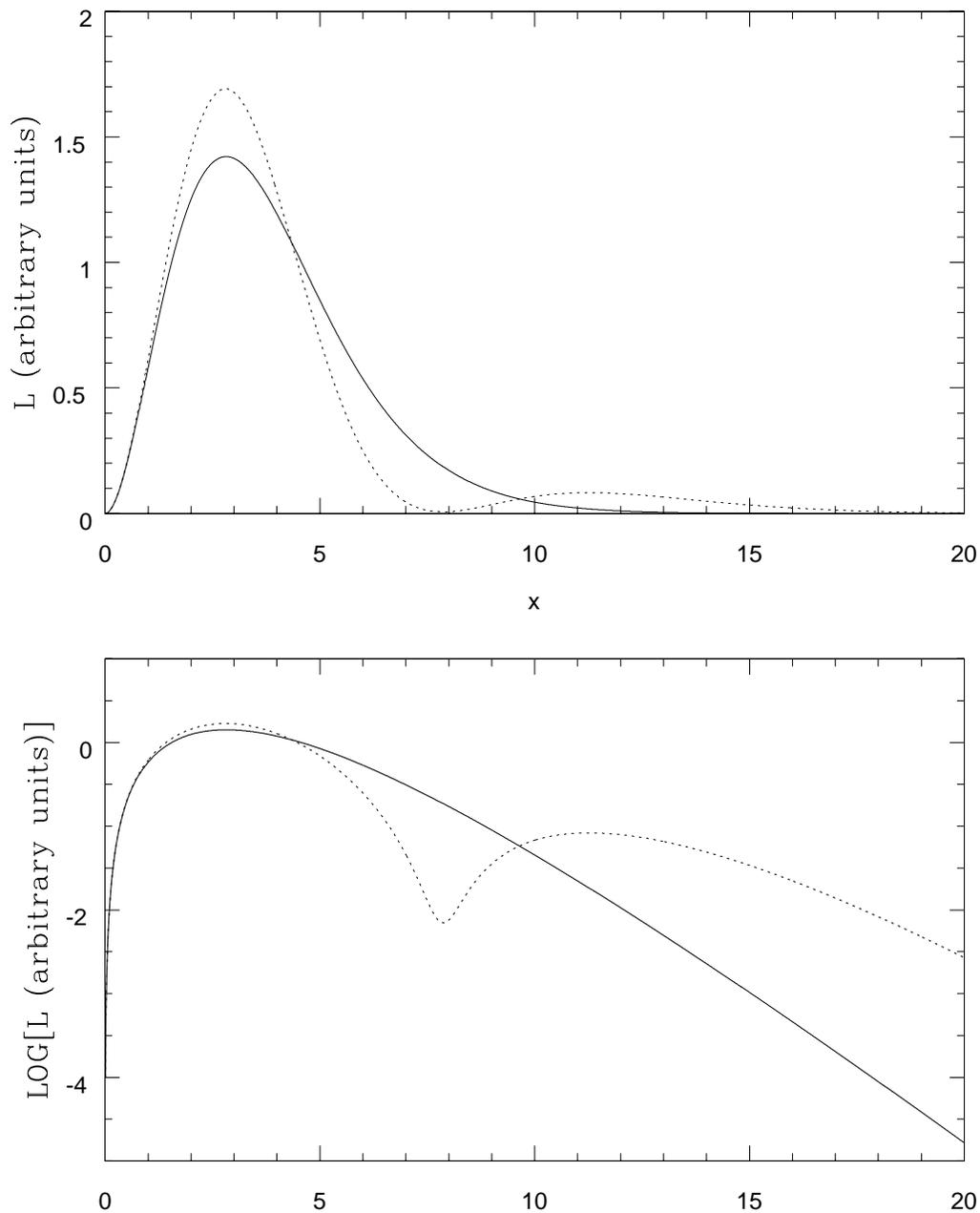}
\caption{In both the upper and lower panels a pure thermal spectrum
(solid curve) is plotted and compared with the perturbed spectrum 
(dotted curve); the case considered is $(\Delta\, t/b)=0.1$. Note that
this value of $(\Delta\, t/b)$ is very close to the maximum value
consistent with the approximation we have made in Equation 8; when the 
value is larger, the
flux becomes negative, hence unphysical,  for a range of frequencies near 
$x=8$.}
\label{fig=fig3}
\end{figure}
\begin{figure}[hbtp]
\leavevmode\epsfxsize=5.6in\epsfbox{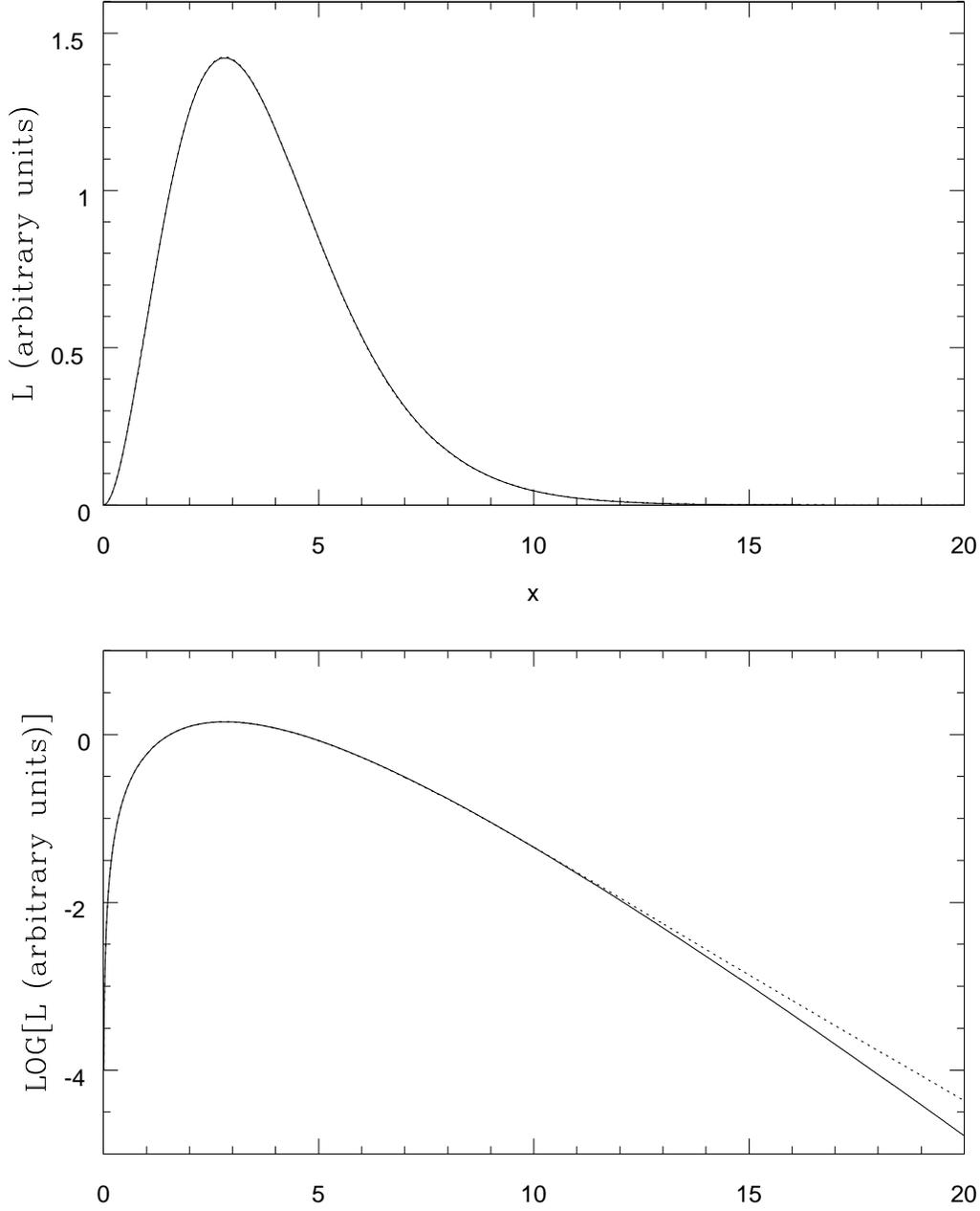}
\caption{In both the upper and lower panels a pure thermal spectrum
(solid curve) is plotted and compared with the perturbed spectrum 
(dotted curve).}
\label{fig=fig3}
\end{figure}
\begin{figure}[hbtp]
\leavevmode\epsfxsize=5.6in\epsfbox{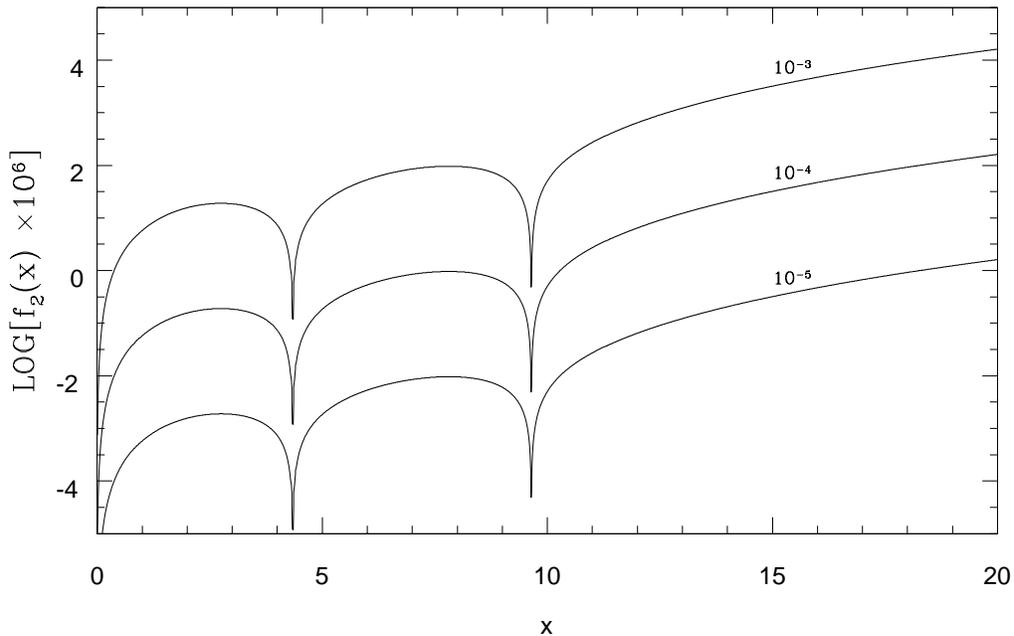}
\vspace{-3.7 true in} 
\caption{Plotted is log$_{10}[f_2(x) \times 10^6]$ vs $x$.
Because $f_2(x)$ (and not $\hat f_2(x)$) is plotted,
each curve is characterized by a value of $(\Delta\, t/b)$;
these values are used to label the curves.
As $(\Delta\, t/b)$ decreases, perturbations due to lightcone
fluctuations become more difficult to discern by eye. This plot illustrates
that precision measurements of the spectral shape can nevertheless be
effective in constraining even small values of $(\Delta\, t/b)$. 
This is particularly so (1) for values of $x$ around extrema
in $\hat f_2(x),$ and (2) at high frequencies, if the thermal spectrum at high frequencies
is not contaminated by contributions due to other physical effects. 
}
\label{fig=fig3}
\end{figure}

\vfil\eject

\section{Application to the Microwave Background}
The spectrum of the microwave background has been measured by the FIRAS
(Far InfraRed Absolute Spectraphotometer) aboard the COBE (Cosmic
Background Explorer) over the range of wavelengths (480-4400 $\mu$m
[$\nu=2-21$~cm$^{-1}$]).  These measurements have constrained departures
from a pure thermal spectrum at a level of a few parts in $10^{-5}$ near
the peak of the spectrum (Fixsen {\it et al.}\ 1996) 
and the absolute temperature
to 2.725$\pm$.002 (95\% CL) (Mather {\it et al.} 1999).
These measurements can be used to constrain the lightcone fluctuations.
\begin{equation}
\Bigg|{{\Delta F(x)}\over{F(x)}}\Bigg| = 
\Bigg|\hat f_2(x)\Bigg| \times \Bigg({{\Delta\, t}\over{b}}\Bigg)^2 < 
\Big| d(x) \Big|, 
\end{equation}
where values of 
d(x) are on the order of $10^{-5}$ near the peak of the spectrum. 

We have used two methods to derive limits on
${{\Delta\, t}\over{b}}$ from the microwave data.
First, we proceeded in exact analogy to
the work carried out by Fixsen {\it et al.} 1996; 
in that paper limits on Comptonization and 
on a chemical potential were derived, based on fits to the 
COBE-measured spectrum. In \S 3.1, limits on the microwave background
are put on the same footing, yielding the
$95\%$ confidence limit ${{\Delta\, t}\over{b}} < 0.0012$. In \S 3.2 we take
another approach, using a point-by-point analysis.
This approach allows us to identify the points that provide 
the strongest constraints, and
show that they produce a result
consistent with the result derived by considering the
full spectrum. Such a point-wise analysis could be used
for other thermal spectra. If, for example, measurements at only
a limited number of points are possible, our approach illustrates
which are the most useful points to sample. Alternatively,
 the point-wise approach could strengthen
the results based on the microwave data if one or 
two high-precision spectral measurements 
of the microwave background can be made
at higher frequencies. 

\subsection{Constraints Based on Spectral Fits}

To derive limits for $(\Delta t/b)^2$ we linearize the distortion in
$(\Delta t/b)^2$ to approximate $\Delta F$ as follows. [See Eqs.\ (8) and (11).]
\begin{equation}
\Delta F(x)=F_0(x)\, \hat{f_2}(x)\ \Big({{\Delta t}\over{b}}\Big)^2,   
\end{equation} 
where $x= 0.5278\, \nu.$ Note that $x,$ which is dimensionless
in units with $c=1$, is defined in Equation (10); here we consider
the specific case of the microwave background, with 
 $T=2.725 K.$

Fixsen {\it et al.} (1996) provide a table with residuals ($r$), 
the $1\, \sigma$ uncertainties
($c$), a galactic template ($G$), and correlations ($Q$). 
Together with the derivative of the Planck function, $dF_0/dT$, we have
all of the components required to fit the quantum gravity fluctuations, 
paralleling the fits appropriate to the other distortions 
(Fixsen {\it et al.} 1996, Eq.\ 3).

The $dF_0/dT$ and galactic spectra are included here (as there) to allow for
uncertainty in the temperature (which was determined from the data) and
the amount of residual galactic contamination.  The full covariance
matrix was used (it can be derived from $\sigma$ and $(\Delta t/b)^2$).  The result of this
fit is $(\Delta t/b)^2=-1.8\times 10^{-7}\pm 7.1\times 10^{-7}$.  Physical measurements 
of $a$ 
should 
yield a non-negative result;  since, however, the negative
result we have found is only at the level of $-.25\sigma,$ it is likely
to represent only statistical errors in the estimation of $a$.

We can then use the derived uncertainty of $a$ to place an upper limit on
$a$ and hence on $\Delta t$.  For a 95\% confidence limit 
\begin{equation} 
(\Delta t/b)^2<2\, (7.1\times10^{-7}),  
\end{equation} 
\begin{equation} 
\Delta t/b<.0012.
\end{equation} 

Figure~7 illustrates this point. Shown are 
the measured residuals from the
analysis of the FIRAS data; a thermal spectrum with
$T=2.725$ K has been subtracted, as has the
dipole contribution and the contribution
likely to be due to our own Galaxy.
 The long-dashed curve corresponds to
 the largest
value of the chemical potential (due to energy input before $z=10^5$)
consistent with the data, and the short-dashed curve corresponds
to the maximum effects of Comptonization (occurring after $z=10^5$)
consistent with the data. 
 as shown in Figure 7.  Since $b=1.76\times 10^{-11}$ for
$T=2.725~$K, this translates to $\Delta t<2.1\times10^{-14}$~s.\footnote{
The
mean frequency over the path of the photon is 
actually $\sim 3$ times the frequency
we measure today, assuming a flat Einstein-de Sitter model. 
This is similar to the result we would have derived in other cosmological
models. Thus, the limits we have derived above are higher by a factor
$\sim {3}$ than those that would have been derived had the 
calculations of Yu \& Ford (1999) been carried out in an expanding spacetime.}

\begin{figure}[hbtp]
\leavevmode\epsfxsize=5.6in\epsfbox{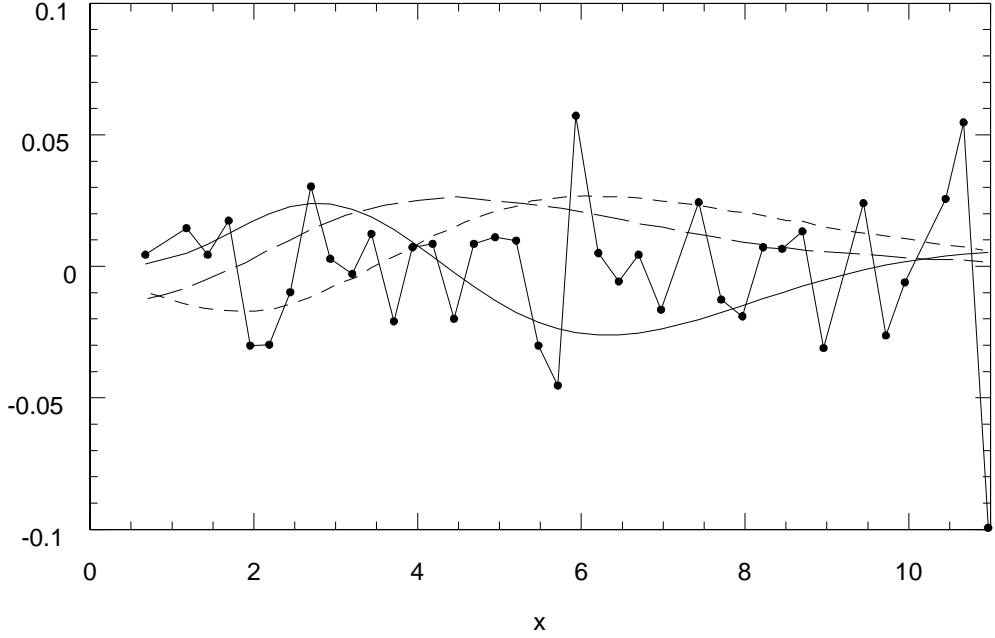}
\vspace{-3.5 true in}
\caption{Deviations from a thermal spectrum.
The unit along the vertical axis is MJy/sr;
the unit along the horizontal axis is the
same dimensionless variable $x$ we have been using throughout this paper.
We have translated from the frequency units of cm$^{-1}$ 
used by
Fixsen {\it et al.} (1996): $x=\nu_{Fixsen}\ (h\, c)/(k\, T)$.    
The jagged line, with data points indicated
by the filled circles, shows the measured residuals from the 
analysis of the FIRAS data; a thermal spectrum with
$T=2.725$ K has been subtracted, as has the 
dipole contribution and the contribution
likely to be due to our own Galaxy.
The solid curve shows the contribution of \lcf s for
$\Delta\, t/b = 0.0012.$ 
  The long-dashed curve corresponds to
  the largest
value of the chemical potential (due to energy input before $z=10^5$)
consistent with the data, and the short-dashed curve corresponds
to the maximum effects of Comptonization (occurring after $z=10^5$)
consistent with the data.}
\end{figure}

\subsection{Constraints Based on Individual Points}     
Each measurement at a given frequency (characterized by 
$x=0.5278\, \nu$),  
provides a constraint on $\Delta\, t/b.$  
\begin{equation}
\log\Bigg[{{\Delta\, t}\over{b}}\Bigg]< -{{1}\over{2}} 
\Bigg\{\log\Bigg[\Big| d(x)\Big| \Bigg] + 
\log\Bigg[\Big| \hat f_2(x) \Big| \Bigg]\Bigg\}.    
\end{equation}
Thus, to derive the tightest limits (i.e., lowest upper bound) 
on $\Delta\, t,$ we want to find the minimum value of
right-hand side of the above equation consistent with the data.

To do this, we considered each frequency, $\nu,$ at which the 
microwave background was measured. 
Let $F(x)$ represent
the best-fit thermal spectrum.  
We estimate $d(x)$ in two ways: $d_1(x)=|{r(x)}|/F(x),$ 
$d_2(x)=c(x)/F(x)$, where $r(x)$ represents the residuals and 
$c(x)$ represents the $1\, \sigma$ uncertainties, respectively. 
Note that $d_1(x)$ and $d_2(x)$ are both 
positive, so we omit the absolute value signs wherever $d(x)$ appears
below.  
For each frequency at which a spectral measurement was made,
we define ${{\Delta\, t}\over{b}}(x)_{ul}$ to be that value of
${{\Delta\, t}\over{b}}$ that turns (16) into an equality.
\begin{equation}
\log\Bigg[{{\Delta\, t}\over{b}}(x)\Bigg]_{ul}= -{{1}\over{2}}
\Bigg\{\log\Bigg[d(x)\Bigg] +
\log\Bigg[\Bigg|\hat f_2(x)\Bigg|\Bigg]\Bigg\}. 
\end{equation}
In this way, each measurement provides [for each way of measuring
$d(x)$] an upper bound on ${{\Delta\, t}\over{b}}.$ 
The lowest value of the
upper bounds so obtained is then the tightest limit that can be
derived from this type of point-by-point analysis of the FIRAS data
set.    
Figure~8 illustrates the results.
The points shown as open circles are $\{d_2(x), {{\Delta\, t}\over{b}}(x)_{ul}\};$
the minimum value of ${{\Delta\, t}\over{b}}(x)$ achieved by
any of these points is $0.0015.$ 
The points shown as crosses are $\{d_1(x), {{\Delta\, t}\over{b}}(x)_{ul}\};$   
the minimum value of ${{\Delta\, t}\over{b}}(x)_{ul}$ achieved by
any of these points is $0.0006.$

The more sophisticated global analysis of the spectrum carried out in
\S 3.1
produces an 
upper  limit that lies between the values obtained by using $d_1$ and $d_2.$ 
This can also be illustrated graphically. If, for example, we
plotted the curves corresponding to $\Delta\, t/b=0.0006$ and
a succession of values up to and including $\Delta\, t/b=0.015$
on this graph, it would be clear that the smaller value is too low and
that the upper limit is too high. In fact, such an analysis
leads to the result:    
\begin{equation}
{{\Delta t}\over{b}} <    0.001^{+0.0003}_{-.0003}.
\end{equation}
Since $b$ (in Equation 10) has the value $1.76 \times 10^{-11}$ s
for $T=2.725$ K, this translates to
\begin{equation}
{\Delta t} < 1.8^{+0.5}_{-0.5} \times 10^{-14}\ s.
\end{equation}
This is consistent with the results of the more sophisticated analysis of
\S 3.1, demonstrating that reliable results can be derived with
an analysis that focuses primarily 
measurements at a few well-chosen frequencies. By ``well-chosen frequencies", 
we mean 
frequencies at which the effects of  lightcone fluctuations are
maximized, as shown in Figures 2 and 6. 
\begin{figure}[hbtp]
\leavevmode\epsfxsize=5.6in\epsfbox{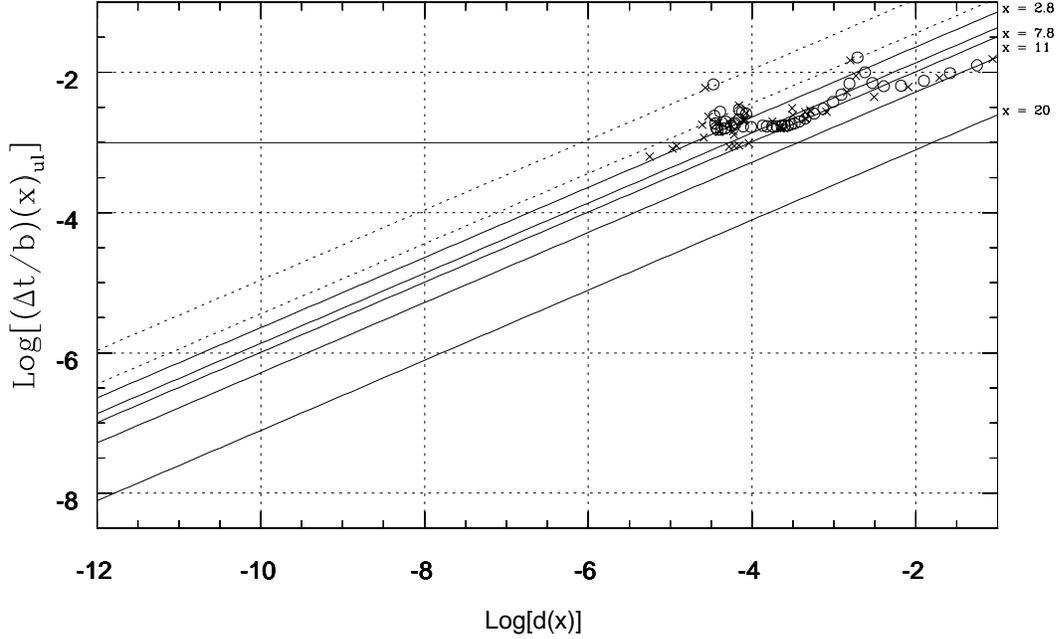}
\vspace{-3.75 true in}
\caption{
The log of the fractional deviation from a pure thermal spectrum
is plotted along the horizontal axis. Along the vertical axis is
plotted 
${{\Delta t}\over{b}}(x)_{ul}$ (i.e., the upper limit allowed
for each value of $x$,  
as defined in Eq.\ (17)). Each point shown corresponds to
the FIRAS measurement of the microwave background at a 
particular frequency, $x.$ 
The points $\{d_1(x), {{\Delta\, t}\over{b}}(x)_{ul}\}$, shown as crosses, are
determined by the residuals; 
the minimum value of ${{\Delta\, t}\over{b}}(x)_{ul}$ achieved by
any of these points is $0.0006.$
The points $\{d_2(x), {{\Delta\, t}\over{b}}(x)_{ul}\}$, shown as open circles,
are determined by the standard deviations; 
the minimum value of ${{\Delta\, t}\over{b}}(x)$ achieved by
any of these points is $0.0015.$
The two cusps with the highest values of ${{\Delta\, t}\over{b}}(x)_{ul}$
correspond to values of $x$ for which $\hat f_2$ is near zero.
The continuous lines each plot Eq.\ (17) for a given value of $x$.
Note that the best limits (i.e., the lowest upper bounds)
 can be obtained for values of $x$ associated
with large values of $|\hat f_2(x)|$. 
This trend is exhibited for the lines with values of $x$ shown
to the right of the panel. Generally, larger
values of $x$ allow stronger constraints to be placed,
although exceptions are illustrated by the
 dotted lines, which correspond to $x\sim 4.3$
(upper line) and $x=\sim 9.6;$ in both cases, $\hat f_2(x)$ is 
approaching zero.   
}
\end{figure}
   
\section{Relating $\Delta t$ to Metric Fluctuations}

We have demonstrated that astrophysical measurements can either measure or place
constraints on the value of $\Delta t$. In this section we demonstrate
how such  
measurements or constraints can provide information about quantum
fluctuations of the spacetime metric.

In a fixed, classical spacetime, the flight time of a pulse traveling from
a source to a detector is uniquely
determined by the spacetime metric and by the path of the pulse.
The flight time will generally differ from what might have been expected
in a flat spacetime.
An example from classical relativity is the time delay
of radar signals passing near the limb
of the sun, which has been used as a test of general relativity.
The fact that the signals
pass through a region with a Schwarzschild metric leads to an increase in
the flight time as compared to what would be expected in flat spacetime.

If, however, the metric were to change, so would the flight time.  
Gravity waves provide examples from classical relativity;
a laser interferometer gravity wave detector can be viewed as measuring
variations in the path length, or equivalently flight time, in the
arms of the interferometer as a gravity wave passes by.
Quantum fluctuations of the metric must also lead to
fluctuations of the flight time. Consider a  
quantum spacetime, represented by an ensemble of classical geometries. 
Each member 
of this ensemble will have its own value for the flight time. The variance 
in flight time obtained when we average over the ensemble is the quantity
 $\Delta t$.
 
Let us suppose that the average spacetime metric is that of Minkowski
spacetime. Further let $r$ be the distance between the source and 
detector as measured in the Minkowski metric. The classical flight 
time of light pulses is just $t = r$, in units where the speed of 
light is unity. The geodesic interval function $\sigma$ is defined 
as one-half of the square of the geodesic distance between a pair of 
spacetime points. Let $\sigma_0 = \frac{1}{2}(t^2-r^2)$  be this
function in flat spacetime. As expected, $\sigma_0 = 0$ when $t=r$. 
Suppose that as a result of a perturbation of the metric, the actual 
flight time becomes $t = r + \delta t$. The geodesic interval function is now
\begin{equation}
\sigma = \frac{1}{2} [(r+\delta t)^2 - r^2] = r \delta t + O((\delta
t)^2) \,.
\end{equation}
  Thus the first order variation in $\sigma$ is $\sigma_1 = r \delta t$.
Suppose that the flight time now fluctuates so that  the average of 
$\delta t$ vanishes, but the average of its square does
not. The root-mean-square fluctuation in flight time is then
\begin{equation}
 \Delta t = \sqrt{\langle (\delta t)^2 \rangle} = 
\frac{\sqrt{\langle \sigma_1^2 \rangle}}{r} \,.
\end{equation}
 
The computation of $\Delta t$ in a particular model of quantum gravity 
effects is reduced to the computation of $\langle \sigma_1^2 \rangle$.
It is shown in Ford (1995) and Ford \& Svaiter (1996) that
\begin{equation}
\langle \sigma_1^2 \rangle = \frac{1}{2} \sigma_0 
\int d\tau_1\, d\tau_2\, u_1^\mu u_1^\nu u_2^\rho u_2^\sigma \,
                     \langle h_{\mu\nu}(x_1) h_{\rho\sigma}(x_2) \rangle \,,
                                 \label{eq:sig1gen}
\end{equation}
where $\tau$ is the affine parameter along and $u^\mu$ is the tangent vector
to the unperturbed geodesic. The above expression is a double integral
of the renormalized graviton two point function, 
$\langle h_{\mu\nu}(x_1) h_{\rho\sigma}(x_2) \rangle$, taken over an 
unperturbed  geodesic. This two point function is the crucial quantity
which a quantum gravity model must provide. 
Because different theories of quantum gravity yield different
two point functions, measurements or constraints on the value of $\Delta t$
can potentially discriminate among the theories. 

For example,
a natural starting point for quantum calculations may be the
graviton two point function of Minkowski
spacetime.
If, however, it is not
renormalized, the integral diverges.
The usual
response to this problem is to require that the renormalized two point
function vanish in Minkowski spacetime. If this is the correct
solution, then $\Delta t = 0$, and there are no lightcone 
fluctuations in Minkowski spacetime.
Many current theories of quantum gravity do not, however,
use 4-dimensional Minkowski spacetime as the starting point.
For example, quantum gravity derived in the context of 
string theory (Ellis, Mavromatos, \& Nanopoulis 2000), 
spacetime foam (Garay 1998), and canonical
quantum gravity in the loop approximation (Gambini \& Pullin 1999)
 all predict
lightcone fluctuations. 

Here we describe  specific models in which lightcone fluctuations are
expected. These models are 
Kaluza-Klein theories with compact extra dimensions. In
these models, our four-dimensional world is a subspace of a higher dimensional
spacetime. The reason that the extra dimensions are not readily observed
would be that they are compactified, so that one can travel only a very
small distance in a higher dimension before returning to the same point.
However, confining a quantum field to a small region enhances its quantum
fluctuations. This effect is a consequence of the uncertainty principle,
and is analogous to what happens to a single quantum particle confined in a
small region of space.
One effect of extra compact space dimensions is to create a spectrum
of massive modes in the theory. This is the ``Kaluza-Klein tower''
of masses predicted by such theories, which are inversely proportional to
the compactification length. Thus as the length becomes smaller, these
masses increase. It is sometimes argued that the effects of
compactification should become smaller in this limit because large masses
should give a smaller contribution to virtual processes. This argument
is incorrect, however. The effect of the infinite sum of masses 
scales oppositely with length from that of a single mass, as can be
understood from the above uncertainty principle argument.
 Thus Kaluza-Klein theories can predict observably
large lightcone fluctuations. The precise prediction of a particular model
can be found only by a difficult calculation, and tend to be very sensitive to
the details of the compactification model. In Yu \& Ford (1999, 1999b, 2000), 
$\Delta t$ was
computed for a variety of models which postulate flat extra dimensions.
In the case of the five dimensional  Kaluza-Klein theory, which has one
extra space dimension of length $L$, it was found that
\begin{equation}
\Delta\, t \approx \left(\frac{r}{L}\right)\, t_{pl} \, ,
 \end{equation}
where $ t_{pl} = 5.4 \times 10^{-44} s$ is the Planck time. 
The value of $r$ depends upon the cosmological model. To derive an
estimate for the limit that the microwave background spectrum places on
the value of $L$, we simply take $r = 10^{28} cm,$ roughly the distance
traveled by a classical light wave in $10$ Gyr. With this estimate,
we find that $L$ must be greater than or equal to approximately a millimeter.
That is
\begin{equation}
L> {{r\, t_{pl}}\over{\Delta\, t_{max}}} = 0.3\, mm\,  
		\Bigg({{r}\over{10^{28}cm}}\Bigg)\,
                                \Bigg({{1.8 \times 10^{-14}}\over{\Delta\, t}}\Bigg)   
\end{equation}
 
Thus, our constraint on $\Delta t$ from the microwave background,
places a very strong constraint on $L$. This constraint
is so strong  
that it essentially rules out the five dimensional  Kaluza-Klein theory.

Note that the form of $\Delta t$ given in Eq. (24) was calculated
in flat spacetime, rather than in a Robertson-Walker spacetime.
In principle, there will a be correction due to the spacetime curvature
in an expanding universe. However, this should be very small because
the wavelength of the photons is always very small compared to the
horizon size. (See footnote 1.)
Apart from this correction, there is the issue
of the interpretation of the length $r$. In flat spacetime, this is
the distance from the source to the detector as measured in the latter's
frame of reference. In the case of the expanding universe, we can
imagine a sequence of comoving observers spaced in such a way that              
spacetime is nearly flat in between each pair of observers. The proper
distance which a light ray travels between a given pair is just the
flight time in the frame of one of the observers, in units where
the speed of light is unity. Thus the net proper distance traveled by the
ray as measured by this sequence is just the elapsed comoving
Robertson-Walker time. Thus our key assumption is that we can use the
flat spacetime result, with $r$ being taken to be this time. Since 
our chosen numerical value for $r$, and the associated limit on $L$,
are basically order of magnitude estimates, changes due to using
a Robertson-Walker metric are within our estimated uncertainties.
Note that, even the weakest result consistent with these
uncertainties still eliminates the Kaluza-Klein model we have considered.

As discussed in \S 5 and noted in Yu \& Ford 2000,
even stronger constraints can, in principle, be obtained
by studying spectra peaked at higher frequencies.
The 
microwave background, however, provides advantages over many other
presently available sources because its spectrum is very
precisely measured, and the light reaching us
has traveled over the longest possible baseline for an 
electromagnetic signal.

There is an additional advantage in using the microwave background: its
relatively low frequency obviates complications that might be related
to the {\it correlation time} for quantum metric fluctuations.
The correlation time, $\tau_c$, can be thought of as being the characteristic
time scale of the metric fluctuations. Recall that what we really measure is not
 the actual flight time of a pulse, but rather the difference in flight times of
a pair of pulses. If the two pulses are emitted within a time of $\tau_c$ of
one another, they essentially propagate through the same classical spacetime
geometry and their arrival times are not affected by the metric fluctuations.
It is only when the pair of pulses are separated by more than $\tau_c$ that
the variation in their arrival times is expected to be of order $\Delta t$.
Thus one can use data on the shapes of spectra to constrain quantum gravity
models only when characteristic frequency of the spectrum is less than about
$1/\tau_c$. As in the case of $\Delta t$, the correlation time $\tau_c$ needs
to be obtained by a difficult calculation in a specific model. 
One might expect $\tau_c$ to be of order $L$, as this is the 
characteristic wavelength of the graviton modes which are most
altered by the compactification. However, in the five dimensional
Kaluza-Klein model, one actually finds (Yu and Ford 2000) the much
smaller value of $\tau_c \approx L^2/r$ when $r \gg L$. It is therefore
possible, at least in principle, for measurements of spectra
peaked at higher frequencies
to also place meaningful constraints on this model. 
The corresponding calculation has not, however,
been performed for other models, so their correlation times are unknown. 
Thus, although data
on high frequency spectra potentially lead to stronger constraints on $\Delta t$,
they can also be limited in their usefulness in constraining models which
predict relatively long correlation times. In this sense, data on lower
frequency spectra, such as the microwave background, provide information
that cannot be obtained from higher frequency spectra.
 
We are naturally also interested in constraints that may apply to other
quantum theories. Interestingly enough, however, the calculations
that predict the size of \lcf s are sensitive to the details
of the quantum theory. For example, Kaluza-Klein models involving more than one
flat extra dimension, at least up through 11-dimensional theories,
 do not lead to significant lightcone fluctuation effects.
Yet, for other theories
spanning the broad range of possible quantum theories of gravity,
including Kaluza-Klein models in which the extra dimensions are not flat, 
it could well be that predictions would have observable
consequences for the microwave background and can therefore be
tested within the framework we have established here.
Indeed, the microwave background would provide
the ideal test for such a theory if the computed correlation time is
significantly longer than that computed for the 5-dimensional 
Kaluza-Klein model with one flat extra dimension.


\section{Summary and Prospects} 

We have pointed out that the best way at present to observe the
effects of the   
minute differences in arrival times predicted by \lcf s
 is through spectral measurements. 
We have computed the perturbation of a thermal spectrum due to 
frequency-independent \lcf s. 

\subsection{Analytic Results}

We find the form of the perturbed spectrum 
to be distinctive. More power is added near the peak of the
thermal spectrum and also at frequencies greater than roughly
$3$ times the frequency at the peak, with a pronounced 
decrease in power just before the rise at high frequencies.
For values of ${{\Delta\, t}\over{b}} > 0.1,$
the lightcone fluctuations play such an important role that
the approximation we have made by terminating the series expansion
at second order breaks down; such large \lcf s would
cause the spectrum to deviate so significantly from
the thermal form that a thermal spectrum alone would not necessarily
be the obvious candidate for a model fit. 

Whatever the actual value of ${{\Delta\, t}\over{b}}$, the
\lcf s place a distinctive stamp on the spectrum. Thus,
if $\Delta\, t$ is large enough to render \lcf s observable,
measurements taken at a range of frequencies should
be able to confirm that deviations from the pure thermal form
are due to \lcf s and not to some other effect.
For ${{\Delta\, t}\over{b}}< 0.1,$ measurements 
either to discover or constrain \lcf s are most effective
when taken in regions around the extrema of $\hat f_2(x)$;
i.e., near $x=2.8$ and $x=7.8$, as well as
at values of $x$ greater than $10.$ Although we have not calculated
the exact result for larger values of ${{\Delta\, t}\over{b}}$, these
same regions of the waveband will almost certainly be the ones in which the
effects of \lcf s are most pronounced.
 
It is important to note that 
the quantity which is constrained by the data is not $\Delta\, t$
itself, but rather the parameter ${{\Delta\, t}\over{b}}.$ Thus,
even 
if the \lcf s are frequency-independent, the size of the parameter
\begin{equation} 
p(T) = {{\Delta\, t}\over{b}}= {{\Delta\, t\ k\, T}\over{h}} 
\end{equation} 
is proportional to the temperature of the
thermal source. Writing $p(T)$ in this way makes it clear that
the largest effects are expected for the hottest sources.
Suppose that we have observed the spectrum from a thermal source
of temperature $T.$  
Let ${\cal D}_{meas}$ represent  a measure of the observational 
limits on fractional deviations from a thermal spectrum.
These observations allow us to place an upper
limit on $\Delta\, t.$
\begin{equation} 
\Delta\, t(T)_{ul} \sim 2.1 \times 10^{-14}\,  s\,  
\Bigg({{{\cal D}_{meas}}\over{10^{-5}}}\Bigg)^{{1}\over{2}}  
\Bigg({{0.0002\, eV}\over{k\, T}}\Bigg),    
\end{equation}
If, for 
example, 
measurements as sensitive as those we now have for the
microwave background were possible
for GeV sources,
the limits on $\Delta\, t$ could be
as small as $10^{-26}$ or $10^{-27}$  s.

We return to the question of whether \lcf s lead to angular effects. 
In the appendix we argue that angular effects should
occur at higher order than the spectral effects we have studied here.
There are two important caveats, however. The first is that the
arguments are developed in the context of a
linearized theory. The second, and perhaps most important,
is that we have assumed translational symmetry, which
does not strictly hold when the photons travel through a curved spacetime. 
This issue must therefore be revisited for each quantum gravity
model under study. In quantum gravity models that differ from the specific
Kaluza-Klein case considered here, a spread in angle due to \lcf s
could influence our inferences about the   
isotropy
of the CMB. 
If the source of radiation is point-like, 
angular spread due to \lcf s could
lead to an observed ``fuzziness" in the image.
   
\subsection{Overview on Testing Theories of Quantum Gravity}

The computations necessary to make specific
predictions for a given quantum theory of gravity are difficult, and
have not yet been carried out for many theories. The general pattern
is that there are $3$ relevant time scales. The first is the 
Planck time, $\tau_{pl} \sim 5.4 \times 10^{-44}$~s. 
The second time scale is the time scale
set by the average difference in arrival times $\Delta\, t.$
The value of $\Delta\, t$ depends on the fundamental underlying
theory and on the path from source to detector. This is also 
true of the third 
time scale, the correlation time, $\tau_c > \tau_{pl} $. 
$\tau_c$
is the characteristic
time scale of the metric fluctuations as integrated along the path 
to the detector. Pulses
received at times
differing from each other by less than $\tau_c$ will
have experienced more-or-less  
the same spacetime as they travel toward the detector,
and so the influence of quantum fluctuations
will not be discernible in effects that rely on differences in
their arrival times. Thus, the detection of quantum fluctuation
effects is possible only if the condition $\Delta\, t > \tau_c$ is
satisfied.

Because both $\tau_c$ and $\Delta\, t$ are model-dependent,
separate computations must be carried out for each theory to
be tested. In addition, both quantities are path dependent, and this
must be taken into account as well.  
In general, the value of $\tau_c$ places a lower bound,
equal to a few times $\tau_c$, for the minimum value of
$\Delta\, t$ that would be detectable. 
If the predicted value of $\Delta\, t$ is larger than this minimum
value, then Eq.\ (26) provides a guide as to the
temperatures of the sources (and the sensitivity of spectral
measurements)  best suited to either measure
the effects of the fluctuations or else falsify the theory.  


\subsection{The Microwave Background}   

We have shown that data taken by FIRAS can, taken by itself,
rule out one quantum theory of gravity, a five-dimensional Kaluza-Klein
model in which the ``extra" dimension is flat. In this theory,
$\Delta\, t$ would have to be larger than allowed by the FIRAS
observations in order for the size of the extra dimension to be
small enough that it should not already have had other observable
consequences. Thus, no further measurements are required to eliminate
this theory from consideration.  

Our analysis of the microwave background data is  
the most detailed analysis to place upper limits on
$\Delta\, t,$ and provides the most secure upper limits to date.  
Furthermore, the correlation times
for   
other models of quantum gravity may tend to be longer,
making it difficult to
rule them out by studying high energy emission. 
Such models could possibly  be ruled out by the
the bound we derived based on the microwave
background.
Further theoretical calculations of $\Delta t$ and of
the correlation time for other theories of quantum gravity will
be required to determine how useful limits derived from the
microwave background will be. 
Since the limits may be important, we next turn to consider
what would be needed to use the microwave background to derive
a lower upper bound on $\Delta\, t.$

\subsubsection{Lowering the Upper Bound on $\Delta\, t$} 

There are two ways future measurements of the
spectrum of the microwave background could lower the upper bound 
on $\Delta\, t$ we have derived.
The first way is by decreasing the uncertainty limits in the wavelength
regime already studied by COBE. 
To improve the limits on $\Delta\, t$ by an order
of magnitude, we would have to decrease the uncertainty limits on the 
measurements by roughly $2$ orders of magnitude. This is because
improvements in constraints on ${{\Delta F(x)}\over{F(x)}}$ by  a factor 
{$I$}, translate to improving the limits on
${{\Delta t}}$ by a factor {$I^{{1}\over{2}}$}.

The second way to lower the upper bound
on $\Delta\, t$ is to recognize that 
the fractional effects of lightcone fluctuations become larger
at higher frequencies (even in the case we have considered in which
$\Delta\, t$ is frequency independent). It would
therefore appear to be considerably
easier to find evidence of them by extending measurements of the CMB
to higher frequencies. Although this is true in principle, 
radiation from the Galaxy and far infrared background become
        more significant at higher frequencies, making it difficult to
to unambiguously attribute radiation at
higher frequencies to the microwave background.
Thus, improved models of Galactic emission are required.
 
Both Figures 6 and 8 demonstrate that decreasing the upper bound of
$\Delta\, t$ by as much as $2$ orders of magnitude
would require either a remarkable increase in the precision
with which we measure the spectrum of the microwave background
and/or a much better understanding of 
Galactic and far infrared
        extra-Galactic emission. 
While significant
improvements of either type are unlikely in the short term, the 
fact that we may still be decades away from a definitive theory
of quantum gravity means that we should consider the long-term prospects as well
as the prospects for the near term.  

\subsection{Identification of Spectral Distortions with Lightcone 
Fluctuations} 

If significant deviations from a thermal spectrum were to be discovered,  
it would naturally be important to consider whether we would be able
to distinguish a signal due to lightcone fluctuations from
signals that might be associated with other effects.
Figure 7 illustrates that the effects of a chemical potential and/or
Comptonization have a spectral pattern
sufficiently different from that predicted by \lcf s, that 
we should be able to distinguish the latter from the
other effects. 

In fact it is interesting to note that, when we consider
the limiting values of parameters consistent with
the COBE data, the effects of Comptonization and the effects of
\lcf s would very nearly cancel.
This has the interesting implication that, on the basis of the
microwave data alone, the combination of 
Comptonization and \lcf s cannot be
ruled out.
Since, however, the physical effects giving rise to 
Comptonization are completely different from those associated
with \lcf s, a nearly exact cancellation between these two effects is
is highly unlikely. Nevertheless, it is interesting to note that
the CMB data can definitively rule out Comptonization effects, 
only if supplemented by additional information, e.g., 
if stronger upper limits can eventually  be
placed on $\Delta\, t$ by studying emission of radiation with
higher energies.


\subsection{Thermal Radiation from Other Sources}

Because $\Delta\, t_{ul}$ depends more strongly on
$T$ than on ${\cal D}_{meas},$ the best way to lower the
upper bound we have derived is to observe the spectra of
higher-temperature sources, even if the precision of the
observations is not as good as that achieved by FIRAS. 
The challenge then is to identify high-temperature 
sources of thermal radiation. Thermal spectra are not the 
norm for X-ray and $\gamma$-ray sources. For example,
the accretion processes that power most X-ray binaries typically
give rise to spectra that are more complicated. Nevertheless
there are suggestions that the X-ray spectra associated with
some processes are thermal. Neutron stars cooling after nuclear-burning
induced X-ray bursts may be one example; the data suggest
that these and perhaps other cooling neutron stars
may be emitting blackbody spectra (see, e.g., Rutledge {\it et al.} 1999). 
There are two problems
however. The first is that present data do not constrain
the spectra very well; i.e., significant departures from
a thermal spectrum may be compatible with existing observations.
The second problem is that the physics of the source systems
is complicated and even a perfectly thermal spectrum can be
distorted by interactions with material near the emitter.
For example, even if a white dwarf or
neutron star emits a thermal spectrum, interaction of the
emergent radiation with the atmosphere of the compact object
alters the spectrum. Basic physical principles can be used
to compute the spectral form of the alterations; 
more work needs to be done, however, to develop 
models good enough that we can 
 subtract atmospheric effects from the observed spectrum
in a way that is as reliable as the subtraction of, e.g., the
dipole contribution to the microwave background.  

Both the challenge of better spectral measurements, and
the challenge of better spectral models will be
met.
While progress in the next few years may not be
significant for the purpose of studying \lcf s, 
enough progress can be expected to occur
over a period of decades.
Since the formulation of a viable and testable 
theory of quantum gravity is likely to take at least
as long as the necessary astrophysical advances,
high-energy  measurements 
should eventually play an important role in
studying \lcf s.

\subsection{Short-Term Prospects}

\noindent{\bf The Microwave Background:\ } 
The microwave background data already rule out the
5-dimensional theory in which the ``extra" dimension is flat.
It is very likely that this data will also place significant 
constraints on other quantum theories of gravity. Progress
awaits computations of $\Delta\, t$ and $\tau_c$ for those
theories.

\noindent{\bf Other Spectra:\  } 
Among thermal spectra,
we can gain the greatest advantage, i.e., place the most stringent
upper bounds on $\Delta\, t,$  by studying the spectra of
high-temperature sources. However, as pointed out above,
present-day limitations
of both our observations and theoretical models    
may not allow thermal spectra to play a major role in
the study of \lcf s in the immediate future. Instead, 
line emission may play the most important role
in the short-term. (See, e.g., data sets such as those described by
Songaila {\it et al.} 1994.) High-energy lines potentially provide the 
strongest constraints.   
For example, by  studying X-ray lines with {\it Chandra}, 
we can potentially place an upper bound on $\Delta\, t$ that is
$\sim 8$ orders of magnitude smaller than the one derived in
this paper, 
 although we may lose $3-4$ orders of magnitude because the
precision of the X-ray measurements now possible is in the
range of $1-10\%$, instead of at the level of a
few parts in $10^5.$ In brief, we can expect X-ray measurements possible in the short-term to reduce the upper limit
on $\Delta\, t$ to $10^{-19}-10^{-18}$ s. If gamma-ray lines can
be used for similar investigations, even smaller spreads in arrival
time can be either measured or constrained.
The caveat that accompanies all such constraints, is that the value
of the correlation time $\tau_c,$ 
computed in the context of the theory we seek to test,
must be smaller than the limits on $\Delta\, t.$ 
 
\subsection{Long-Term Prospects}

The quest for a quantum theory of gravity has been one of the major themes
of physics in the twentieth century. As the search continues into the twenty first  
century, uncertainties about important features of the theory abound.
Consider an optimistic scenario in which, during the coming decades, 
theorists manage to hone in on a small range of 
candidate theories that make clear and verifiable predictions.
Even in this case, the fact that key predictions 
may not be testable in earth-based accelerators  
will undoubtedly mean that astrophysical observations during the
present century and beyond will play
important roles in eliminating some theories and possibly verifying others.

We therefore want to emphasize that, in order to assess the role
astrophysical observations might play in studying \lcf s, we must
take the long view. It is true that the spectra of high-temperature sources
are not yet well constrained by the observations, and that  
many physical effects we now understand imperfectly (such as the influence of
neutron-star atmospheres) can themselves alter spectra that
start out as purely thermal.
In $10$ years, $50$ years, $100$ years, however, we will have made improved
measurements with much lower levels of uncertainty than at present.
Furthermore, theoretical models of even complex phenomena such as
neutron star atmospheres, and the Galactic contribution
to radiation across the spectrum, will be well-developed
and tested. Observers  will therefore eventually be able to  
subtract from their observed spectra effects associated with many 
complicated physical situations (e.g.,  local conditions near the emitter),
with as much confidence as
the FIRAS team was able to subtract out the dipole contribution
from their observed spectrum.  
This, together with a wealth of well-studied sources at
different temperatures and lying at different distances
away from us,  
means that a long-term ambitious program like the one described below
will be possible. 

First, a succession of measurements at higher and higher energies
can place constraints on the level of \lcf s, eliminating 
some candidate theories of quantum gravity. If, at some energy,
definitive evidence of \lcf s is found, then investigations of
sources at similar temperatures but at different distances will
help to constrain the form of metric fluctuations. (Note that
sources lying behind large mass distributions also provide
ways to sample different path lengths.) 
Investigations of 
sources at similar  distances but at different temperatures
will help to explore possible frequency dependence of \lcf s.

In short, we think that during the coming decades, radiation from a wealth of astrophysical
sources will provide even more stringent constraints on \lcf s,
or else will discover definitive evidence of their existence.
In either case, these astrophysical systems will teach us much more
about \lcf s and
the quantum gravity theories that predict them. 

\bigskip
\bigskip
\bigskip
\bigskip
\centerline{\bf Acknowledements}

We would like to thank A. Loeb for encouragement and
he, P. Kaaret, and M. White for discussions. 
We are grateful to Zahra Valimahomed for help with the figures.
 This work was supported in part by NSF through grant PHY-9800965. 
 
{}  

\bigskip
\bigskip

\centerline{\bf Appendix: Deviations as a Function of Angle} 

\def\be{\begin{equation}}
\def\ee{\end{equation}}	
\def\ben{\begin{eqnarray}}
\def\een{\end{eqnarray}}


The question of angular spreading is not central to the results
   derived in the body of this paper. 
This is because we are interested
   in the FIRAS temperature determination and not in the variation
   of temperature across the sky. Thus, even if there were significant
   angular fluctuations,
the effect would still
   be unobservable, since the scattering is from another region of CMBR
   at almost the same temperature.

Angular dependence is potentially
   interesting, however, because it provides another effect which
   can be used to test quantum gravity. We therefore
wish to discuss whether the same effects which produce quantum lightcone
fluctuations, in the form of a time delay or advance, $\Delta t$, can also
produce a significant angular deflection, $\Delta \theta$.

Consider the  geodesic equation:
\be
{d^2x^{\mu}\over d\lambda^2}=-\Gamma_{\alpha \beta}^{\mu}{d x^{\alpha}\over d\lambda } 
{d x^{\beta}\over d\lambda }=-\Gamma^{\mu}_{VV}\,.
\ee
Here we have introduced null coordinates $V=x+t$ and $U=x-t$ and taken $V$ to be the 
affine parameter on a $U=const$ geodesic. Hence, in the linearized theory of gravity, one 
has 
\be
{d^2y\over dV^2}=-\Gamma^y_{VV}\approx-{1\over 2}\eta^{yy}(2h_{yV,V}-h_{VV,y})\,.
\ee
Integrate this equation to get
\be
{dy\over dV}=\int_0^{V_0}\,{d^2y\over dV^2}\,dV=-\bigl[h_{yV}\bigr]^{V=V_0}_{V=0}
+{1\over 2}\int_0^{V_0}\,h_{VV,y}\,dV\,.
\ee
The first term here is a boundary term which we want to discard. This can be accomplished, 
for example,  by requiring that the perturbation vanishes at the endpoints of the geodesic. 
Thus 
\be
{dy\over dV}={1\over 2}\int_0^{V_0}\,h_{VV,y}\,dV\,.
\ee
The angular deflection is given by
\be
\Delta \theta ={dy/dV\over dx/dV}=\int^{V_0}_0\, h_{VV,y}\,dV + O(h^2)\,.
\ee
 Here we have appealed to the fact that 
\be
x={1\over 2}(V+U)\,,
\ee
and
\be
{dx\over dV}={1\over 2}
\ee
to the lowest order in $h$. Therefore, the mean squared angular deflection due to quantum 
metric fluctuations can be expressed as 
\ben
\langle (\Delta \theta)^2\rangle &=& \int_0^{V_0}\,dV\,\int_0^{V_0}\,dV'\,
\langle h_{VV,y}(x) h_{VV,y'}(x')\rangle\nonumber\\
&=& \int_0^{V_0}\,dV\,\int_0^{V_0}\,dV'\,\partial_y\partial_y'
\langle h_{VV}(x) h_{VV}(x')\rangle\nonumber\\
&=&\partial_y\partial_y'\int_0^{V_0}\,dV\,\int_0^{V_0}\,dV'\,
\langle h_{VV}(x) h_{VV}(x')\rangle\nonumber\\
&=&0\,.
\een
Here we have assumed that $\langle h_{VV}(x) h_{VV}(x')\rangle$ is a constant independent
of $y$ and $y'$ along the geodesic, since $\langle h_{VV}(x) h_{VV}(x')\rangle$ is only a function 
of $y-y'$ because of the translational symmetry and $y=y'$ along the geodesic to the first order in 
$h$.
There are no angular fluctuations in the order we are working 
here.

Note that our derivation here stricly applies only for a flat background
spacetime, not the Robertson-Walker spacetime one is presumably dealing
with in cosmology. In particular, our assumption that the graviton two
point function is tranlationally invariant may not hold in a more general
spacetime. However, one still expects the angular fluctuations to be
unobservably small, based upon the following argument. We have just shown
that on a flat background, the angular flucutations vanish in an order in
which the lightcone fluctuations (fluctuations in flight time) can be
nonzero. Thus when the angular flucutations are nonzero, we expect them to
be smaller than the fluctuations in flight time. Specifically, this
implies $\Delta \theta = \Delta y/r \ll \Delta t/r$. However, $\Delta t/r$
is the fractional variation in the flight path, which is vastly smaller
than the fractional spread in frequency due to lightcone fluctuations,
which is $\Delta t/\lambda$, where $\lambda$ is the wavelength. Thus,
we conclude that variations in frequency due to metric flucuations will
be many orders of magnitude larger than the angular variations.

\end{document}